# Controlled edge dependent stacking of $WS_2$-$WS_2$ Homo- and $WS_2$-$WSe_2$ Hetero-structures: A Computational Study

Kamalika Ghatak[1], Kyung Nam Kang[2], Eui-Hyeok Yang[2,*], Dibakar Datta[1,*]

[1] Department of Mechanical and Industrial Engineering, New Jersey Institute of Technology, Newark, NJ 07103, USA

[2] Department of Mechanical Engineering, Stevens Institute of Technology, Hoboken, NJ 07030, USA

* Corresponding author (Email: eyang@stevens.edu, dibakar.datta@njit.edu)

## Abstract

Transition Metal Dichalcogenides (TMDs) are one of the most studied two-dimensional materials in the last 5-10 years due to their extremely interesting layer dependent properties. Despite the presence of vast research work on TMDs, the complex relation between the electro-chemical and physical properties make them the subject of further research. Our main objective is to provide a better insight into the electronic structure of TMDs. This will help us better understand the stability of the bilayer post growth homo/hetero products based on the various edge-termination, and different stacking of the two layers. In this regard, two Tungsten (W) based non-periodic chalcogenide flakes (sulfides and selenides) were considered. An in-depth analysis of their different edge termination and stacking arrangement was performed via Density Functional Theory method using VASP software. Our finding indicates the preference of chalcogenide (*c*-) terminated structures over the metal (*m*-) terminated structures for both homo and heterobilayers, and thus strongly suggests the nonexistence of the *m*-terminated TMDs bilayer products.

## 1. Introduction:

The discovery of graphene has motivated researchers to find or synthesize similar novel two-dimensional (2D) structures for various practical applications. The non-bandgap nature associated with graphene hinders the implementation of 2D geometry in certain areas such as transistors, sensors, etc.[1] In this regard, various other 2D materials, including graphene, oxide, and chalcogen family, were explored and investigated after the discovery of graphene in 2004.[2] Among the new generation 2D materials, Transition Metal Dichalcogenides (TMDs) are one of the most versatile materials owing to their wide variety of physicochemical, electrical, and mechanical properties that lie in between the semiconductor and metal.[3-5] Moreover, the graphite-like features of the bulk-TMD crystals and possible graphene-like exfoliation of monolayer-TMDs, as well as the semi-honeycomb features, make them an attractive candidate. TMDs have the general formula $MX_2$ (M = Mo and W; X = S, Se, and Te). The crystal structure of TMDs is three atoms thin, where one metal atom is sandwiched in between two chalcogen atoms (X-M-X) via strong covalent bonds (see Fig. 1b). These monolayer 2D TMDs are attached through weak van der Waals forces in the multilayered 3D crystal structures. Among various 2D TMDs, Molybdenum/Tungsten Sulfide/Selenides (Mo/W $S_2$/$Se_2$)[6] are known to be stable at ambient conditions and thus are known to be useful for energy-storage, sensing, electronic and photonic devices applications.

Among the monolayer (ML) TMDs, the direct band gap of $MoS_2$, $WS_2$, and $WSe_2$ fall in the visible to the near-IR range and are therefore well explored for their significant electronic, optical and photocatalytic properties.[3,7-11] ML $MoS_2$ is well studied[12-18] for its transistor applications, excellent carrier mobility, as well as its potential applications in spintronics, valleytronics, bio, and gas sensing applications. On the other hand, $WS_2$ possesses similar crystal structure as of $MoS_2$ and, shows higher quantum efficiency, wider valence band maximum splitting and, lower effective mass of the carrier due to the presence of heavier W atom as compared to the Mo atom.[19-23] However, the synthesis of these ML 2D materials is the first step towards its practical application. The initial step begins with the exfoliation of ML $WS_2$ from the bulk crystal structures. Liquid exfoliation technique by Coleman et.al[24] and the mechanical exfoliation method by Zhang et.al[25] are the two most significant contributions towards the ML $WS_2$ synthesis. Among various chemical techniques, lithium intercalation in $WS_2$ powder, followed by the easy chemical exfoliation of the MLs, is also noteworthy. [26,27]

Apart from the exfoliation methods, various other growth techniques such as Physical Vapor Deposition (PVD)[28-30] and Chemical Vapor Deposition (CVD)[31,32] techniques were adopted. Various CVD techniques using different substrates such as, on Si/$SiO_2$, Ti/$TiO_2$, Graphite, Graphene oxides, Sapphire, Au foil, *h*-BN etc. were also reported.[33-39] Among several existing growth techniques, epitaxial CVD growth mechanism is the most preferred one due to its several technological advantages[40-42] such as reduction of defect density, consistency in the overall growth, growth products with sharper interfaces, the perseverance of in-plane electrical conductivity via the introduction of mirror twin grain boundaries, and simultaneous reduction of tilt grain boundaries.[43-45] However, these experimental growth techniques require further control over the layer thickness, edge sharpness, size, and the quality of the as-grown crystals in order to achieve specific applicability.[33] Along with the ML TMDs, synthesis of van der Waals 2D homo and heterostructures are also becoming an attractive field of research due to their layer dependent vast tunable properties.[6,46,47] Therefore, it is essential to study the stable layered growth products and perform a thorough computational analysis in order to better understand the electronic and associated properties. In this respect, Density Functional Theory (DFT) study can provide deep insights of the edge and stacking dependency of the stable post growth products.

In this work, we have performed DFT calculations to get an in-depth understanding of the structural and electronic properties of homo/hetero bilayers of triangular $WS_2$, and $WSe_2$ flakes. We investigated the nature and strength of the interlayer interactions, by testing various combinations of AA and AB stacking. The most stable and unstable combinations were sorted based on their stacking energy and structural changes along with their electronic distribution. This comparative study also represents the important differences based on homo ($WS_2$-$WS_2$) and hetero bilayers ($WS_2$-$WSe_2$). This study aims to provide a crucial insight based on the existence of a specific combination of growth products towards the growth dynamics. Most of the existing studies on TMD heterostructures are based on periodic structures.[48-51] This work considers non-periodic structures, which are most commonly observed in experiments.[52-54] Our results provide insight into the stability of the post-growth TMD homobilayers and heterobilayers.

## 2. Computational Methodology and Models

### 2.1 Methodology

All the electronic structure calculations were performed using the DFT method as implemented in the Vienna Ab initio Simulation Package (VASP).[55] Projector augmented wave (PAW) pseudopotential is taken for the inert core electrons, and valence electrons are represented by plane-wave basis set.[56,57] The generalized gradient approximation (GGA), with the Perdew–Becke–Ernzerhoff (PBE) exchange–correlation functionals, is taken into account.[58] In order to accurately estimate weak van der Waals interaction, a vdW-correction approach is used. The vdW-density functional (vdW-DF) combines nonlocal correlations directly within a DFT functional. All DFT based calculations in this study have been performed using the optPBE functional within the vdW-DFT family, implemented in the VASP package by Klimes and co-workers.[59-61] The plane wave basis was set up with a kinetic energy cutoff of 400 eV. Due to the non-periodic nature of the structure, the Brillouin zone is sampled using the gamma-$k$-point grid for each mono and bilayers. The box size was kept big enough in order to prevent the periodic image overlap. The distance between two periodic images was maintained to be more than 8 Å on each direction (details are present in the Section S1 of *supporting information (SI)*). The initial distances (W-W interlayer) between the two layers were kept in between the ~6.25−6.30 Å following the previous studies.[62,63] The convergence criterion for the electronic relaxation was kept $10^{-5}$ eV/cell, and the total energy was calculated with the linear tetrahedron method with Blochl corrections. All the internal coordinates were relaxed using conjugate gradient methods until the Hellmann−Feynman forces are less than 0.02 eV/Å. In this study, all the atoms present in the systems were fully relaxed, and no atomic positions were fixed during the optimization method.

The free energy for stacking (stacking energy) was calculated using the following formula

$$\Delta G_f = G_{BL} - (G_{ML} + G_{ML'}) \qquad (1)$$

Where, $\Delta G_f$ signifies the stacking energy of the bilayer TMD, and $G_{BL}, G_{ML}$ or $G_{ML'}$ represent the total free energies of the bilayers and different monolayers. In case of homo-bilayers, *ML* and *ML'* represent the same mono-layers, but they represent different monolayers for the hetero-bilayers.

## 2.2 Models

Our study considers only zigzag edged TMD structures due to their well-known energetic and as well as various electronic preferences over the armchair ones.[64-72] In this regard, triangular MX$_2$ flakes are known to be zigzag edged from each side (see Fig. 1c), and they can either be transition metal terminated (*m-*) or chalcogen terminated (*c-*). More importantly, triangular flakes are one of the well-known geometries formed by the MLs.[73-75] Fig. 1a represents the top view of the hexagonal honeycomb geometry of an MX$_2$ flake with both armchair and zigzag edges. Top and bottom side of the figure represents the armchair edges, while, the left and right sides are the representative of the differently terminated zigzag edges. To understand the influence of the different edge termination towards the bilayer formation, one has to begin the analysis by considering either fully metal or chalcogen terminated MX$_2$ monolayer. In this study, we have investigated the stacking of triangular ML flakes and have considered both the homo and the hetero-bilayers cases. In the case of homo-bilayers, we have considered WS$_2$-WS$_2$. For WS$_2$-WSe$_2$ hetero-bilayer, we have changed the chalcogen atoms (X) of one of the layers to *Se* from *S*. Apart from the edge termination, we have also considered the possible contribution of different stacking of this monolayer MX$_2$ flakes towards the formation of the bilayers. In order to do so, we have only focused on the two most important stacking possibilities vis a vis. AA and AB (see Fig. 2). Our investigation is focused on the zigzag edged WS$_2$ and WSe$_2$ monolayers (ML) with two different edge terminations - a) *m*-terminated (Fig. 1c1) and b) chalcogen (*c*) terminated (Fig. 1c2). Each of these monolayers (MLs) gets stacked together, forming bilayers (BLs) via two different stacking orientations named AA stacked (*m-m* and *c-c*), and AB stacked (*m-m* and *c-c*) as shown in Fig. 2. Here Fig. 2 is a general representation of all the homo/hetero layered (WS$_2$-WS$_2$ (homo) and WS$_2$-WSe$_2$ (hetero)) structures. AA is when the M and M (similarly X and X) sit on top of each other (Fig. 2a). On the other hand, AB represents the situation when M and X (similarly X and M) sit on top of each other (see Fig. 2b). It is important to mention here that due to the high computational cost and non-periodic nature of the models, small triangular flakes were considered to gain insight about the stacking phenomenon. Furthermore, the stacking of the triangular ML flakes was considered in order to fully understand the influence of edge termination of both layers and experimental reports on the stacking of triangular MLs were also present in the literature.[76,77] Since the main purpose is to understand the edge effect, we haven't considered the edge termination/passivation.

## 3. Results and Discussion

### 3.1 Differently Stacked Homo/Hetero BLs Formation of WX$_2$:

After performing a rigorous scaling process (details are given in the section S1 of the *SI*), we have optimized the following formula for the *c*- and *m*- terminated MLs: a) $W_{10}S_{30}$ (*c*-), b) $W_{10}S_{15}Se_{15}$ (*c*-), c) $W_{15}S_{20}$ (*m*-) and, d) $W_{15}S_{10}Se_{10}$ (*m*-). In this manner, we have the following formula for the BLs: $W_{20}S_{60}$ (homo-BL-*c*-AA and homo-BL-*c*-AB), $W_{20}S_{30}Se_{30}$ (hetero-BL-*c*-AA and hetero-BL-*c*-AB), $W_{30}S_{40}$ (homo-BL-*m*-AA and homo-BL-*m*-AB) and, $W_{30}S_{20}Se_{20}$ (hetero-BL-*m*-AA and hetero-BL-*m*-AB). We have relaxed the systems mentioned above to find stable orientations. Fig. 3 represents the optimized structures for all the BLs, and Fig. S1 of the *SI* represents the optimized structures of the above mentioned MLs. It is visually significant from Fig. 3 that for both the AA and AB stacked homo/hetero BLs, *c*-terminated structures are less distorted as compared to their *m*-terminated counterparts. However, it is quite evident from Fig. 3g and 3h that both the *m*-terminated AA/AB stacked heterobilayers are almost distorted from their original trigonal geometry.

To predict the stability of the afore-mentioned BL geometries, we have calculated the stacking energies ($\Delta G_f$). Moreover, a proper structural analysis was also conducted before studying their electronic properties. Table 1 and 2 depicts all the stacking energies values for the homo and hetero BLs respectively. High negative stacking energies of all the *c*-terminated BLs are also indicative of their extreme stability. On the other hand, low negative stacking energies of the *m*-terminated homo/hetero BLs strongly suggests the lower probability of their formation during the growth process. Most importantly, a moderately high positive stacking energy (Table 2) of the *m*-terminated AA stacked $WS_2$-$WSe_2$ BL implies the non-existence of the same during the growth process. We also found that the *c*-terminated heterobilayers are more stable from their homo counterparts since they have higher negative stacking energies.

Structural analyses of these BLs and a comparative study with MLs is the first step to address the differences in the $\Delta G_f$ values. Table 3 (*c*-terminated) and 4 (*m*-terminated) represents the average bond distances of all the BL and ML geometries. We have first considered the average interlayer W-W distances to get an insight of the same towards the stacking energy. In the case of *c*-terminated AA stacked BLs, both the homo and hetero W-W interlayer distances fall under the same range (~6.25-6.28 Å), which is comparable to the previously reported studies[62,63] as well (see Table 3). In the case of AB stacked structures, the W-W interlayer distances increase up to ~6.45 Å since the interlayer W's are not on top of each other due to their stacking arrangement. On the other hand, all the *m*-terminated geometries show increased W-W distances, and they are in the range of 6.56 – 6.84 Å (see Table 4).

It is important to note here that the increased interlayer W-W distances also correspond to their lower stacking energies as compared to the *c*-terminated geometries. Most importantly, the W-W distances are highest in the case of the AA stacked *m*-terminated heterobilayers. The increase in W-W distance weakens the van der Waals bond, which further results in lesser stability. The average intralayer W-W, W-S, and W-Se distances also play important roles in their corresponding energies. It is evident from Table 3 that all the intralayer bond distances (W-S/W-Se/W-W) are almost similar in the *c*-terminated BLs and MLs. Moreover, BLs are achieving the extra van der Waals attraction between the two layers and thus providing extreme stability. On the contrary, the intralayer W-S/W-Se distances of the *m*-terminated BLs vary a lot from their corresponding MLs and increased up to ~0.2 – 0.4 Å (see Table 4). The increase in the intralayer distance indicates the weakening of the bonds, which is in accordance with our findings. Furthermore, it is also evident from the optimized *m*-terminated MLs, that the intralayer bond length distribution varies a lot. A few bonds are shortened, and a few are elongated, resulting in a non-uniform distribution for the same. However, the interlayer bond distances are uniform in cases of all the *c*-terminated MLs and BLs.

### 3.2 Bader Charge Analysis of the BLs:

Bader Charge Analysis[78] is one of the well-known approximation schemes to calculate the total electronic charge around each atom within the Bader volume. We have used post analysis scripts by Henkelman group[79] to calculate the total valence electron distribution around each atom in the system. In our case, the system contains W, S, and Se, and according to our used pseudopotential, each of these atoms contain 6 valence electrons. Thus, the charge is calculated based on a total of 6 electrons. All the *c*- and *m*-terminated bilayers were considered. Electron partitioning is tabulated with their corresponding numbers in section S3 of the *SI*. The net charge on atom can be calculated as follows: a) if an atom contains less than 6 electron and the number of electrons is *x* (*x* can be a fraction too), then that atom is positively charged and the corresponding charge is **+(6-*x*)** and, b) if an atom contains more than 6 electrons and the number of electrons is *y* (*y* can be a fraction too), then that atom is negatively charged and the corresponding charge is **-(*y*-6)**. Fig. S2 and S3 of the *SI* represent the side views of the *c*- and *m*-terminated homo/heterobilayers respectively. All atoms are numbered (this number corresponds to the numbered atoms in the section S3 of the *SI*) and differently oriented such that most of them are visible. From the tabulated electron distribution (Section S3 of the *SI*), it is clear that all the W atoms are positively charged and all the X (S or Se) atoms are negatively charged. However, the distribution of electrons varies in a large extent for each of these systems. The variation is significant in between the *m*- and *c*-terminated systems for the homo and heterobilayers.

The highest and lowest charges on the W, S, and Se were calculated and marked with blue color inside the bracket (Section S3 of *SI*). The AA and AB stacked homobilayers show a uniform electron distribution for each W and S atoms, and it ranges in between ~3-4 electrons for W and ~6-7 electrons for S. Average electron partitions on W and S are in the range of ~3.7 and, ~6.8 respectively. On the other hand, the electron distributions show some amount of non-uniformity for the *c*-terminated AA and AB stacked heterobilayers. W attached to the S atoms show an average electron distribution in the range of ~3.7 and the W attached to the Se atoms show an average electron distribution in the range of ~4.1. As a result, the difference in electron distribution also significant in the S (higher electron distribution) and Se (lower electron distribution) atoms respectively. From our observation, it is clear that the difference in the electron distribution in heterobilayers (presence of two types of W atoms attached to different chalcogen (S/Se) atoms) as compared to the homobilayers acts as a stabilizing factor for the same. Our results suggest that the difference in electron distribution (charge separation) in between the W and X (S/Se) atoms is lower for the cases of *m*-termination and thus account for their lower negative stacking energies.

### 3.3 Charge Density Difference (CDD) Analysis of the BLs:

To study the underlying reason behind the difference in stability of the considered BLs, charge density distribution (CDD) analyses were performed, and the spatial distribution of the electron cloud (isosurface) was plotted. Fig 4 and 5 describe the top and side views of the CDD of the homo and hetero bilayers respectively. The homogeneous charge density distribution for the *c*-terminated homobilayers (see Fig. 4a1 and 4b1) and considerable charge density overlap between the bilayers (see Fig. 4a2 and 4b2) are in accordance with their negative stacking energy. On the other hand, the diffused electron cloud distribution in the triangular edges of the *m*-terminated homobilayers (see Fig. 4c1 and 4d1) and lesser overlap between the bilayers (see Fig. 4c2 and 4d2) strongly support their lesser negative stacking energies as compared to their *c*-terminated analogues.

However, the *c*-terminated heterobilayers show a strong electron cloud overlap between the two MLs (Fig. 5a2 and 5b2) and the extent of charge distribution in both the AA and AB stacked heterobilayers are stronger as compared to the *c*-terminated AA and AB stacked homobilayers. These findings also corroborate their corresponding stacking energy from Section 3.1 as well. Nonetheless, Fig. 5d2 is suggestive of very less electron cloud overlap in between *m*-terminated AB stacked heterobilayer and therefore, validates its lower stacking energy. Most importantly, the *m*-terminated AA stacked heterobilayer (see Fig. 5c2) does not have any charge density overlap in between the two layers, and finally turns out to be an unstable system (positive stacking energy).

**3.4 Density of States (DOS) Analysis of the BLs:**

To understand the changes in the electronic properties of these differently terminated bilayers, we have included the DOS analyses. The atomic DOS analyses give us an insight into the electron density distribution in or around the Fermi level, and the extent of atomic orbital overlap for each of these systems. Fig. 6 and 7 describe the partial density of states (PDOS) plot of the homo/heterobilayers. The AA/AB stacked *c*-terminated homo/heterobilayers (Fig. 6a, 6b, 7a and, 7b) show a little discontinuity in PDOS near the Fermi level ($E_F$) indicating semiconducting nature. However, all the *m*-terminated cases (Fig. 6c, 6d, 7c and, 7d) show continuous PDOS starting from valence to conduction level and thus strongly suggests the metallic nature. Furthermore, all the *c*-terminated homo/heterobilayers show significant overlap between the chalcogen's *p* orbitals and metal's *d* orbital below Fermi level as compared to their *m*-terminated counterparts and thus strongly suggests their extensive stability. In the case of heterobilayers, the extent of *p-d* overlap is much more significant and thus advocate towards their high stability.

From our analyses of the structural and electronic property, and the overall energetics, it clear that both *c*-terminated homo/hetero BLs are more likely to exist as bilayer 2D TMD growth products. The absence of structural deformity, homogeneous charge, and electron cloud distribution combine to form the most stable *c*-terminated geometries. Among *m*-terminated homo BLs, both the AA and AB stacking seems to be stable and are likely to exist, although, they are less stable than their *c*-terminated analogues. On the contrary, the AA stacked *m*-terminated BL is most likely to be non-existent. The instability or the lesser stability of the m-terminated cases can be due to the presence of dangling bonds (unsatisfied valence) of the heavy transition metal W. Due to W-termination, their valency is not satisfied fully and thus leading towards the structural instability. Our study explores three different aspects such as a) the effect of edge termination, b) the effect of different stacking, and c) the comparison of homo and hetero bilayers and draws an insightful picture for the probability of their existence as a bilayer TMD growth product. This study gives a motivation towards the controlled growth process of the homo and hetero layered TMDs and aims to provide a clear path towards the application-dependent TMD synthesis.

## 4. Conclusion

In this DFT based study, electronic insights of the controlled bilayer TMD growth products with the changing parameters were studied. Three main parameters were considered such as different edge termination (*c*- and *m*- terminated), different bilayer stacking (AA and AB), and the change in the chalcogen atoms in one of the layers (homo and hetero). Tungsten (W) based chalcogenides (S and Se) were chosen due to their increasing demand in electronic applications over the well-studied Mo based systems. Any applications of these 2D Tungsten Chalcogenides ($WX_2$) require a well-established controlled synthesis path due to their structural dependent applications. Formation of bilayer flakes of these chalcogenides is a major research area and our calculation suggests the extreme stability of the *c*-terminated cases over the *m*-terminated bilayer growth products for both the homo and heterobilayers. The experimental detection method of presence/absence of particular growth products and their particular ratio in the reaction vessel is very difficult and sometimes results in erroneous prediction. However, this benchmarking computational study draws a clear picture in terms of the bilayer growth products based on the energetics and electronics to achieve an application based controlled bilayer TMD products. Although this study only gives an assessment based on the three most well contributed parameters, more parameters can also be added, and it is currently under investigation.

**Acknowledgement**

This work was supported in part by National Science Foundation award (ECCS-1104870), the Defense University Research Instrumentation Program (FA9550-11-1-0272), and the NJIT faculty start-up grant. We are grateful to the High- Performance Computing (HPC) facilities managed by Academic and Research Computing Systems (ARCS) in the Department of Information Services and Technology (IST) of the New Jersey Institute of Technology (NJIT). Some computations were performed on Kong.njit.edu HPC cluster, managed by ARCS. We acknowledge the support of the Extreme Science and Engineering Discovery Environment (XSEDE) for providing us their computational facilities (Start Up Allocation – DMR170065 & Research Allocation – DMR180013).

## Data Availability

The datasets generated during and/or analyzed during the current study are available from the corresponding author on reasonable request.

## Competing Interests

The authors declare that there are no competing interests.

## Author Contribution

All authors contributed to designing the project. KG performed all the calculations and discussed results with DD, KK, and EH. KG wrote the manuscript. All authors checked the final draft.

# Tables

**Controlled edge dependent stacking of WS$_2$-WS$_2$ Homo- and WS$_2$-WSe$_2$ Hetero-structures: A Computational Study**

Kamalika Ghatak[1], Kyung Nam Kang[2], Eui-Hyeok Yang*[2], Dibakar Datta*[1]

[1] Department of Mechanical and Industrial Engineering, New Jersey Institute of Technology, Newark, NJ 07103, USA

[2] Department of Mechanical Engineering, Stevens Institute of Technology, Hoboken, NJ 07030, USA

* Corresponding author

Table 1. Stacking energies of different edged and differently stacked homobilayers.

| Species | $\Delta G_f$ (eV) |
|---|---|
| AA stacked *c*-terminated homobilayer | -1.8076437 |
| AB stacked *c*-terminated homobilayer | -1.92999697 |
| AA stacked *m*-terminated homobilayer | -0.3802691 |
| AA stacked *m*-terminated homobilayer | -0.1542134 |

Table 2. Stacking energies of different edged and differently stacked heterobilayers.

| Species | $\Delta G_f$ (eV) |
|---|---|
| AA stacked *c*-terminated heterobilayer | -2.64495617 |
| AB stacked *c*-terminated heterobilayer | -1.90643926 |
| AA stacked *m*-terminated heterobilayer | 0.83085255 |
| AB stacked *m*-terminated heterobilayer | -0.31549407 |

Table 3. Average bond distances of intralayer W-S, W-W, and interlayer W-W of *c*-terminated MLs and BLs.

| Species | Avg. W-S (Å) | Avg. W-Se (Å) | Avg. Intra W-W (Å) | Avg. Inter W-W (Å) |
|---|---|---|---|---|
| $WS_2$ *c*-terminated | 2.413018571 | -- | 3.161553333 | -- |
| $WSe_2$ *c*-terminated | -- | 2.613228571 | 3.244215 | -- |
| $WS_2$-$WS_2$ AA stacked *c*-terminated | 2.416548889 | -- | 3.153309091 | 6.28111875 |
| $WS_2$-$WS_2$ AB stacked *c*-terminated | 2.404327778 | -- | 3.137748182 | 6.455065 |
| $WS_2$-$WSe_2$ AA stacked *c*-terminated | 2.416011667 | 2.509388333 | 3.188602727 | 6.25176625 |
| $WS_2$-$WSe_2$ AB stacked *c*-terminated | 2.414475 | 2.52706 | 3.171012727 | 6.45871125 |

Table 4. Average bond distances of intralayer W-S, W-W, and interlayer W-W of *m*-terminated MLs and BLs.

| Species | Avg. W-S (Å) | Avg. W-Se (Å) | Avg. Intra W-W (Å) | Avg. Inter W-W (Å) |
|---|---|---|---|---|
| $WS_2$ *m*-terminated | 2.412232857 | -- | 2.906161667 | -- |
| $WSe_2$ *m*-terminated | -- | 2.50215 | 2.978058333 | -- |
| $WS_2$-$WS_2$ AA stacked *m*-terminated | 2.390242222 | -- | 2.870407273 | 6.58096125 |
| $WS_2$-$WS_2$ AB stacked *m*-terminated | 2.380716667 | -- | 2.896436364 | 6.56902375 |
| $WS_2$-$WSe_2$ AA stacked *m*-terminated | 2.507915 | 2.95827 | 3.186575 | 6.8431425 |
| $WS_2$-$WSe_2$ AB stacked *m*-terminated | 2.519025 | 2.61946 | 3.136821667 | 6.6746925 |

# Figures

**Controlled edge dependent stacking of WS$_2$-WS$_2$ Homo- and WS$_2$-WSe$_2$ Hetero-structures: A Computational Study**

Kamalika Ghatak[1], Kyung Nam Kang[2], Eui-Hyeok Yang[2],*, Dibakar Datta[1],*

[1] Department of Mechanical and Industrial Engineering, New Jersey Institute of Technology, Newark, NJ 07103, USA

[2] Department of Mechanical Engineering, Stevens Institute of Technology, Hoboken, NJ 07030, USA

* Corresponding author

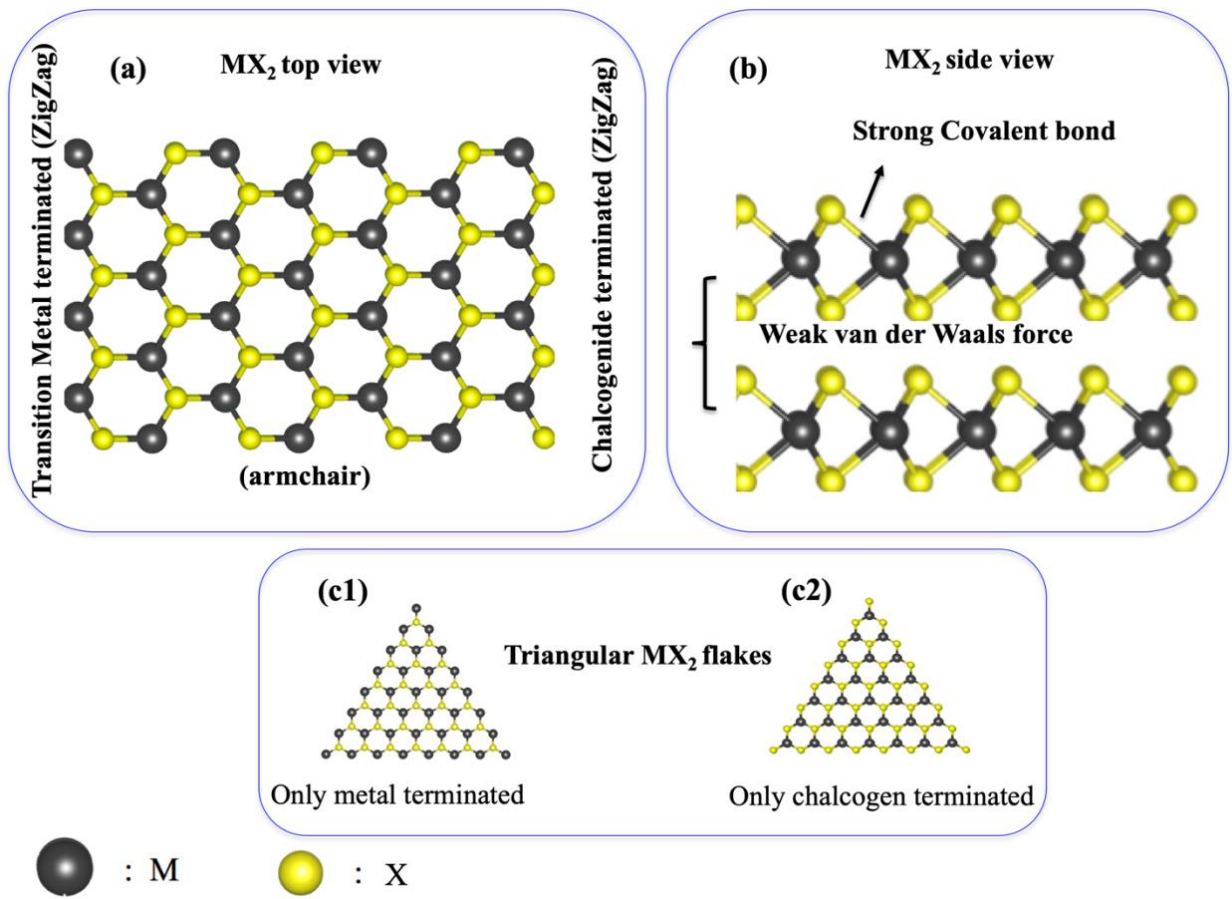

Figure 1. General structural features of 2D-MX$_2$ (a) Top view of 2D-MX$_2$, (b) side view of a bilayer 2D-MX$_2$, (c) triangular MX$_2$ metal/chalcogen terminated zigzag edged structures.

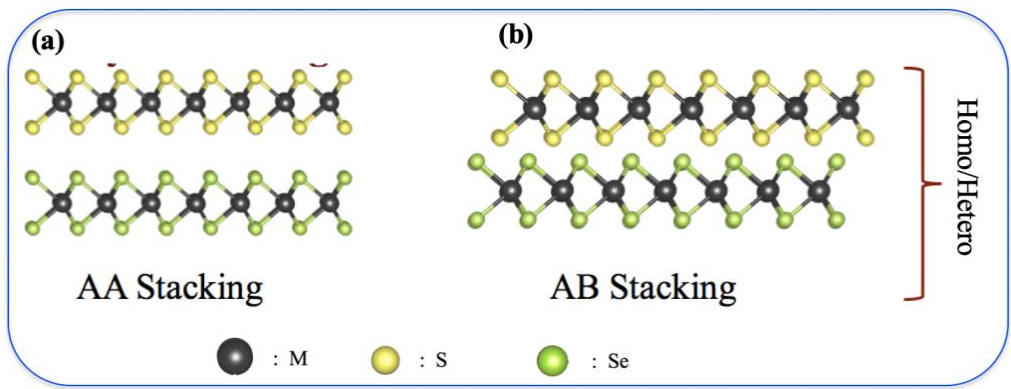

Figure 2. General structural features of 2D-stacked MX$_2$ bilayers (a) AA stacking, (b) AB stacking.

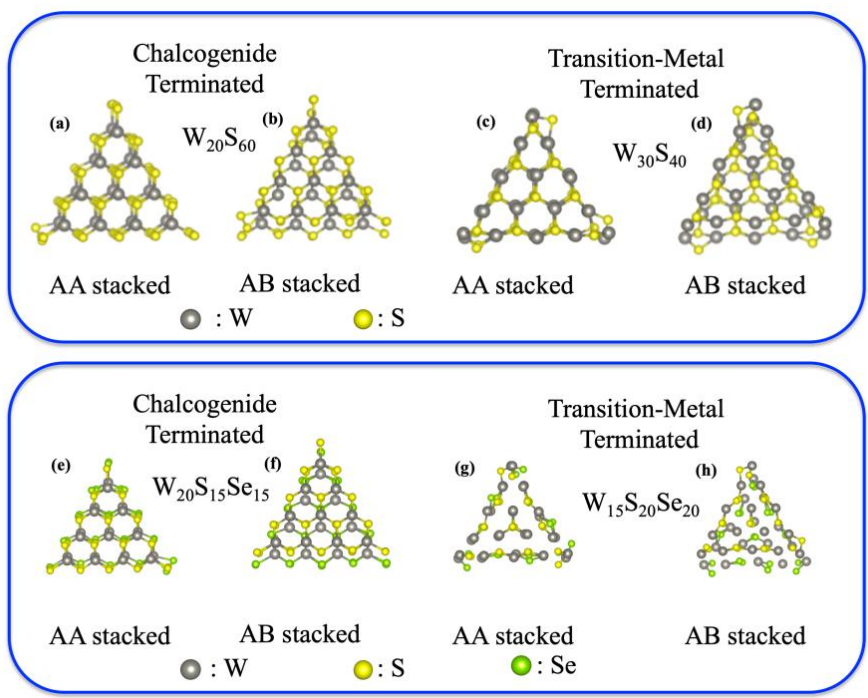

Figure 3. Optimized structures of 2D-stacked MX$_2$ bilayers (a) *c*-terminated AA stacked homo bilayer, (b) *c*-terminated AB stacked homo bilayer, (c) *m*-terminated AA stacked homo bilayer, (d) *m*-terminated AB stacked homo bilayer, (e) *c*-terminated AA stacked hetero bilayer, (f) *c*-terminated AB stacked hetero bilayer, (g) *m*-terminated AA stacked hetero bilayer, (h) *m*-terminated AB stacked hetero bilayer.

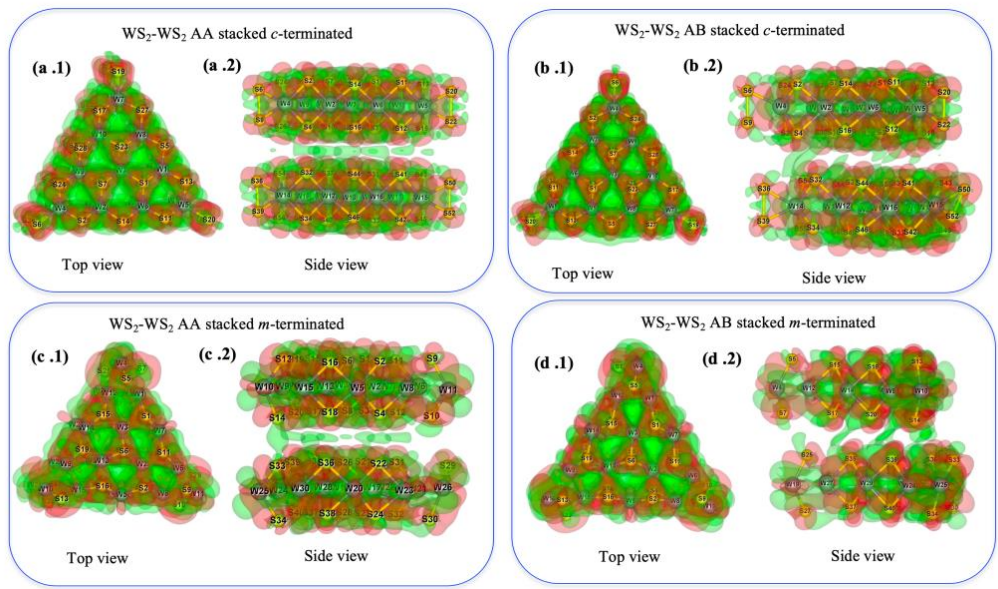

Figure 4. CDD plot of 2D-stacked MX$_2$ bilayers (a) *c*-terminated AA stacked homobilayer, (b) *c*-terminated AB stacked homobilayer, (c) *m*-terminated AA stacked homobilayer, (d) *m*-terminated AB stacked homobilayer.

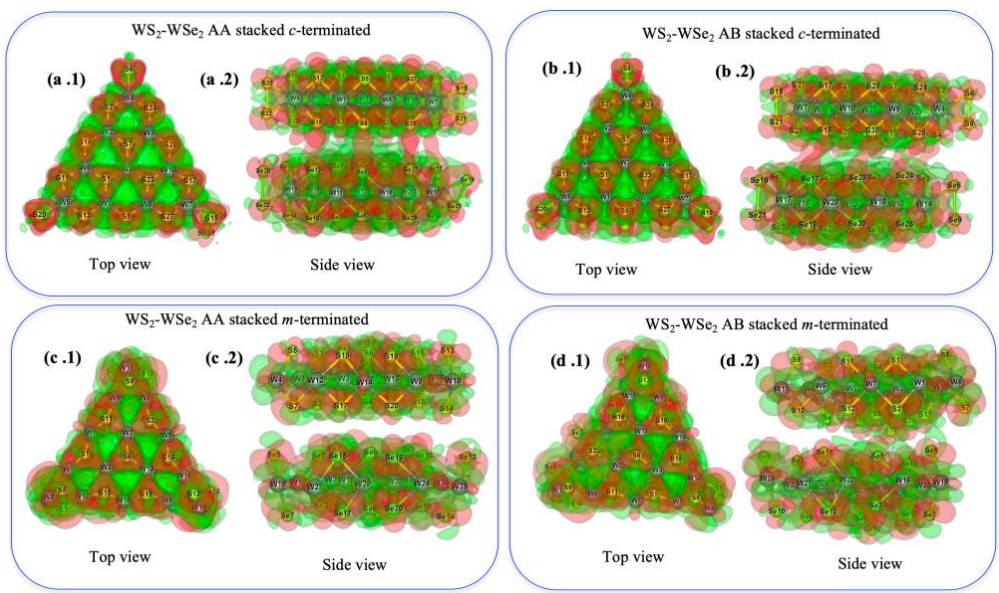

Figure 5. CDD plot of 2D-stacked $MX_2$ bilayers (a) *c*-terminated AA stacked heterobilayer, (b) *c*-terminated AB stacked heterobilayer, (c) *m*-terminated AA stacked heterobilayer, (d) *m*-terminated AB stacked heterobilayer.

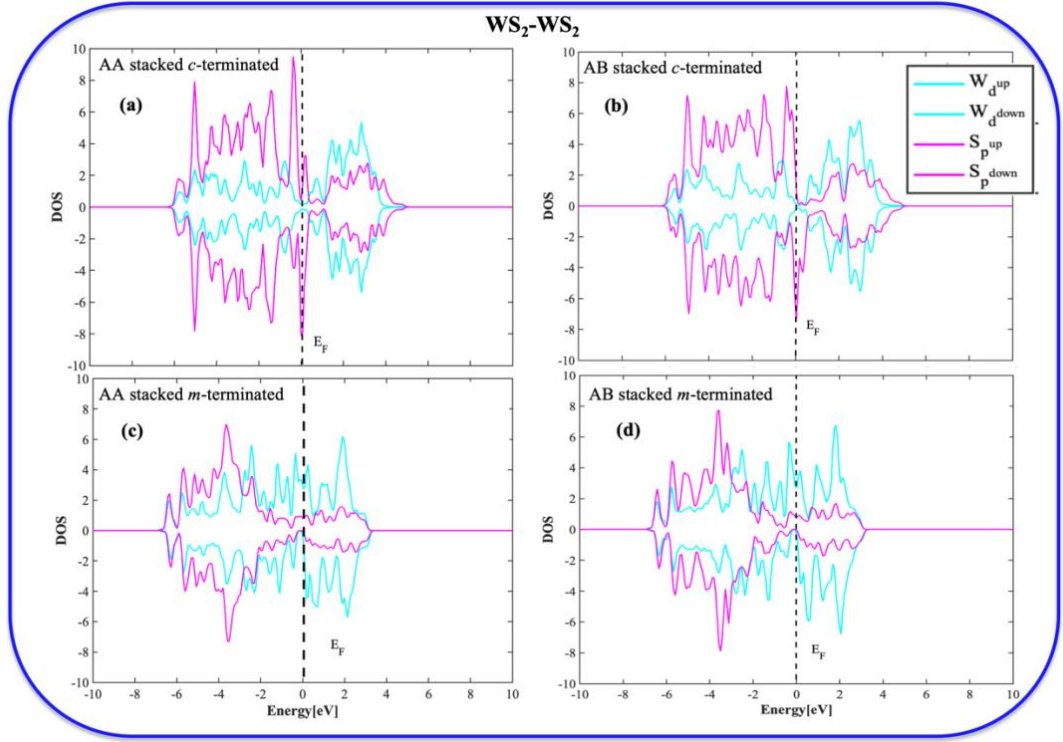

Figure 6. Partial Density of States (PDOS) plots of (a) *c*-terminated AA stacked homobilayer, (b) *c*-terminated AB stacked homobilayer, (c) *m*-terminated AA stacked homobilayer, (d) *m*-terminated AB stacked homobilayer.

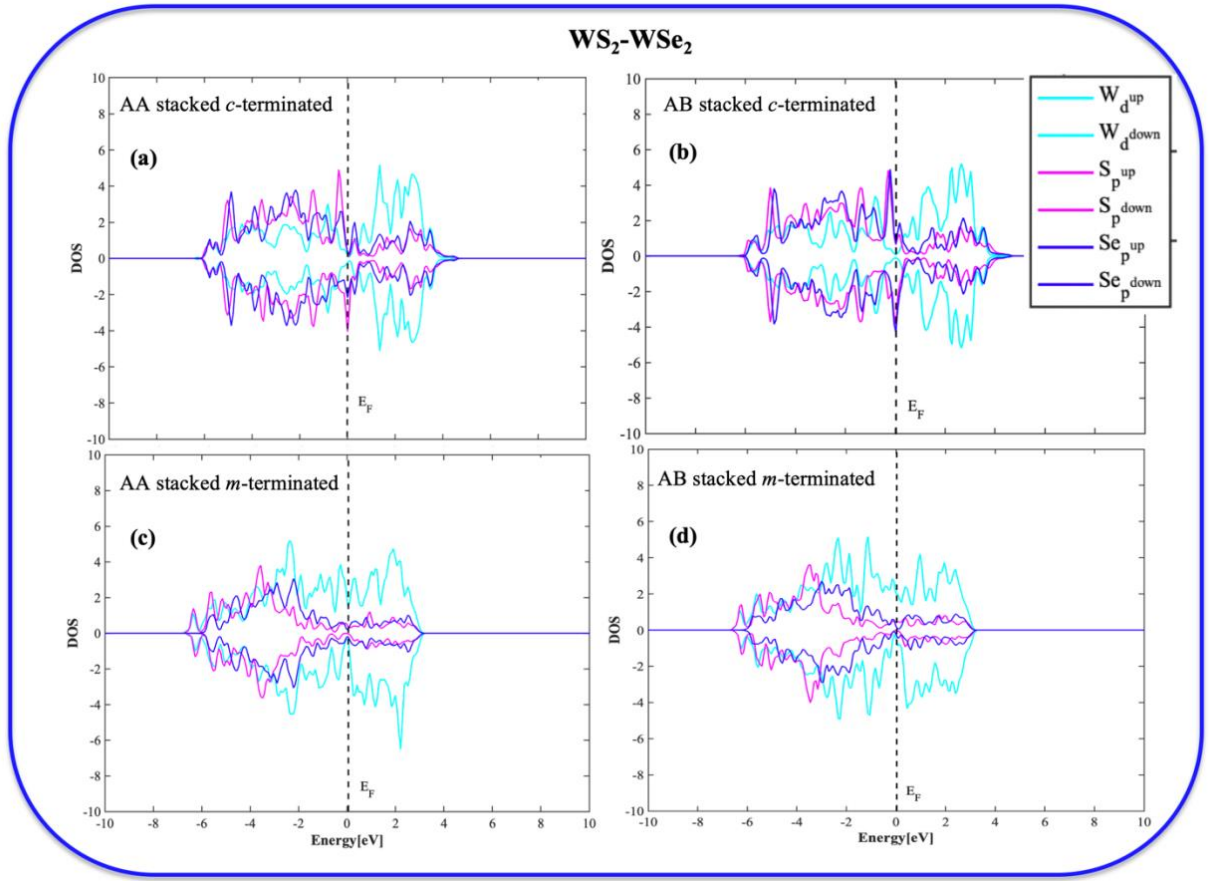

Figure 7. Partial Density of States (PDOS) plots of (a) *c*-terminated AA stacked heterobilayer, (b) *c*-terminated AB stacked heterobilayer, (c) *m*-terminated AA stacked heterobilayer, (d) *m*-terminated AB stacked heterobilayer.

# Supporting Information

## Controlled edge dependent stacking of $WS_2$-$WS_2$ Homo- and $WS_2$-$WSe_2$ Hetero-structures: A Computational Study

Kamalika Ghatak[1], Kyung Nam Kang[2], Eui-Hyeok Yang[2,*], Dibakar Datta[1,*]

[1] Department of Mechanical and Industrial Engineering, New Jersey Institute of Technology, Newark, NJ 07103, USA

[2] Department of Mechanical Engineering, Stevens Institute of Technology, Hoboken, NJ 07030, USA

* Corresponding author

**Section S1. Scaling of the calculation setup (VASP):**

All our structures are non-periodic in nature, and therefore, it requires the increased simulation box size to avoid the error due to the periodic image overlap. We have utilized GPU facility of our institutional cluster KONG for VASP simulation. To scale the kinetic energy cut-off, we have considered one case with a total of 80 atoms (BL-*c*-AA-$W_{20}S_{60}$). For this case, a box size of (35× 35×35) Å with 4 combinations of energy cut-off starting from 300-600 eV (300, 400, 500, and 600 eV) were taken. We have performed 4 single point calculation, and the variation in energy values from 400 eV – 600 eV is insignificant. Therefore, we have considered 400 eV kinetic energy cut-off for all our computations. Moreover, this single point calculation (one geometric iteration) takes time in between ~2-3 hrs. and the time increases with the increasing kinetic cut-off values. These number of geometric iterations for all these structures vary in between ~200-400 geometric iterations depending upon the systems. There are few reasons behind this significant time requirement for all these geometries: a) the presence of heavy transition metal such as W, b) the huge box size, c) high kinetic energy cut-off and, finally d) the non-periodic nature of the calculation. Therefore, an optimized combination of all these issues need to be taken care of in order to resist the huge computational cost. In this regard, we first considered 2 combinations of the cubic cell sizes Comb-A (35 × 35×35) Å$^3$ and, Comb-B (27× 27×27) Å$^3$ and ran a single point calculation to trace the time consumption (row 1 and 2 of Table 1). In these two cases, the average wall time/ iteration is huge, and it will take a lot of time in order to get the full optimization. Therefore, after several trials and error, we have optimized the simulation box size to be Comb-C (24× 22×20) Å$^3$ (row 3 of Table S1), where the distance of the two consecutive periodic images were kept more than 8 Å. The initial lengths of the triangular *c*-terminated and *m*-terminated flakes are ~13.2 Å, and ~11.8 Å.

**Table S1.** Cell size parametrization for TMD BL optimization.

| Species | Stacking | Energy Cutoff | Cell size | Per iteration average time in hrs. |
|---|---|---|---|---|
| BL c-$W_{20}S_{60}$ (Single point) | AA | 400 | 35× 35×35 (comb-A) | 2.10 |
| BL c-$W_{20}S_{60}$ (Single point) | AA | 400 | 27× 27×27 (comb-B) | 1.29 |
| BL c-$W_{20}S_{60}$ (Optimized) | AA | 400 | 24× 22×20 (comb-C) | 0.31 |

**Section S2. Labelled optimized geometries of the ML & BL TMDs:**

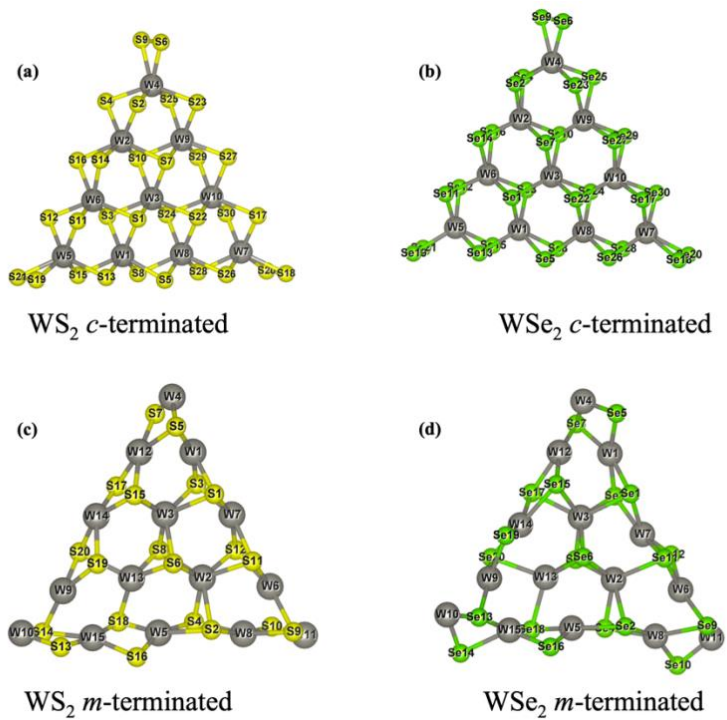

**Fig. S1.** Optimized labelled top views of a) *c*- terminated WS$_2$, b) *c*-terminated WSe$_2$, c) *m*-terminated WS$_2$ and, d) *m*-terminated WSe$_2$.

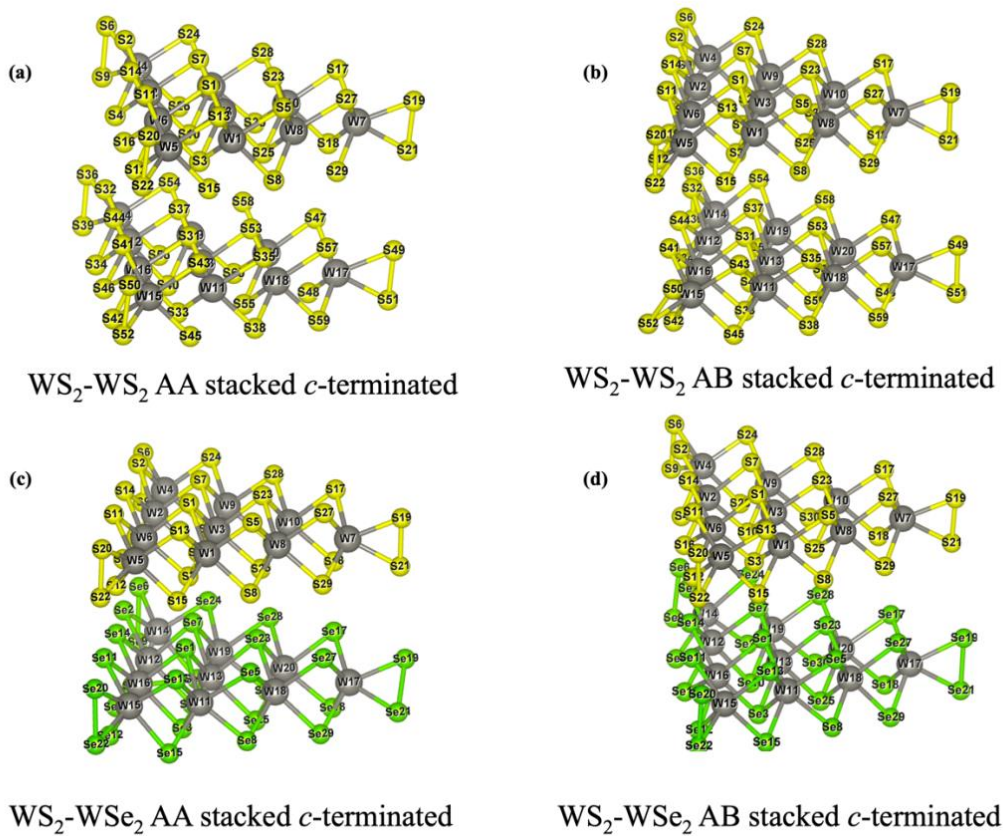

**Fig. S2.** Optimized labelled side views of a) *c*- terminated AA stacked $WS_2$-$WS_2$, b) *c*- terminated AB stacked $WS_2$-$WS_2$, c) *c*- terminated AA stacked $WS_2$-$WSe_2$ and, d) *c*- terminated AB stacked $WS_2$-$WSe_2$.

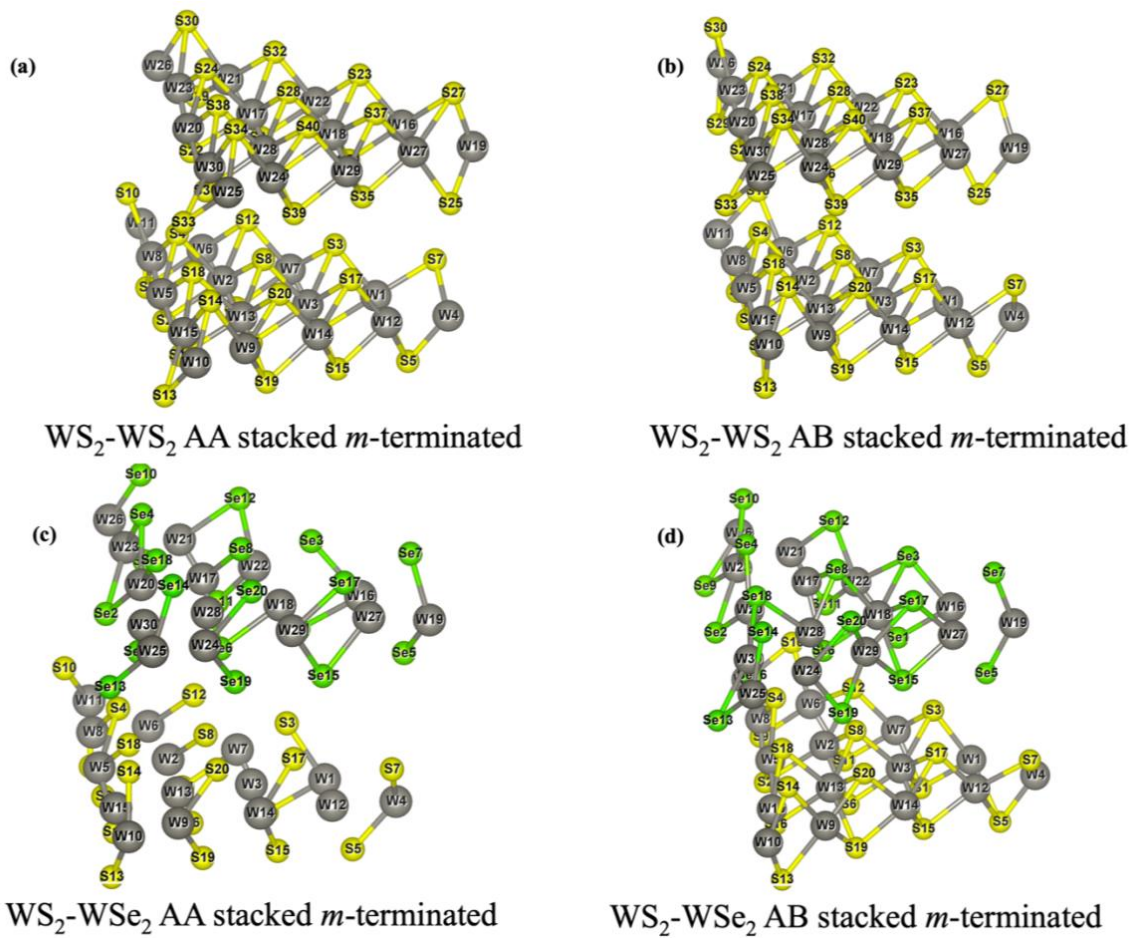

Fig. S3. Optimized labelled side views of a) *m*- terminated AA stacked $WS_2$-$WS_2$, b) *m*- terminated AB stacked $WS_2$-$WS_2$, c) *m*- terminated AA stacked $WS_2$-$WSe_2$ and, d) *m*- terminated AB stacked $WS_2$-$WSe_2$.

**Section S3. Bader Charge Analysis for the corresponding systems:**

All the charges were tabulated based on the atoms number provided on Fig S2 and S3. The highest and lowest positively charged atom's electron partition were given in **Red** color and the negatively charged highest and lowest atom's electron partition were provided in **Green** color. Their corresponding charges were provided inside the bracket and in **Blue** color.

**WS2-WS2 AA stacked c-terminated**

| Atom number | Bader charge |
|---|---|
| W_1 | 3.7922 |
| W_2 | 3.7589 |
| W_3 | 3.9923 |
| W_4 | 3.4615 |
| W_5 | 3.5019 |
| W_6 | 3.758 |
| W_7 | 3.5093 |
| W_8 | 3.8072 |
| W_9 | 3.7804 |
| W_10 | 3.7658 |
| W_11 | 3.7746 |
| W_12 | 3.8198 |
| W_13 | **4.0273 (+1.9727 e)** |
| W_14 | 3.4712 |
| W_15 | 3.5118 |
| W_16 | 3.804 |
| W_17 | **3.2494 (+2.7506 e)** |
| W_18 | 3.7994 |
| W_19 | 3.7856 |
| W_20 | 3.7683 |
| S_1 | 6.9879 |
| S_2 | 6.8927 |
| S_3 | 7.0206 |
| S_4 | 6.8634 |
| S_5 | 6.8692 |
| S_6 | 6.3651 |
| S_7 | 6.9586 |
| S_8 | 6.7752 |
| S_9 | 6.3498 |
| S_10 | 7.0261 |
| S_11 | 6.8733 |
| S_12 | 6.8514 |
| S_13 | 6.8569 |
| S_14 | 6.839 |
| S_15 | 6.7738 |
| S_16 | 6.728 |
| S_17 | 6.8421 |
| S_18 | 6.8119 |
| S_19 | **6.2582 (-0.2582 e)** |
| S_20 | 6.2695 |
| S_21 | 6.4046 |
| S_22 | 6.4253 |
| S_23 | 6.9701 |
| S_24 | 6.8541 |
| S_25 | 7.0187 |
| S_26 | 6.7725 |
| S_27 | 6.8647 |
| S_28 | 6.8109 |
| S_29 | 6.7748 |
| S_30 | 6.7844 |
| S_31 | 7.0063 |
| S_32 | 6.8741 |
| S_33 | 6.9732 |
| S_34 | 6.8698 |
| S_35 | 6.7764 |
| S_36 | 6.3423 |
| S_37 | 7.0006 |
| S_38 | 6.8332 |
| S_39 | 6.3697 |
| S_40 | 6.9707 |
| S_41 | 6.8096 |
| S_42 | 6.8324 |
| S_43 | 6.7561 |
| S_44 | 6.7076 |
| S_45 | 6.8631 |
| S_46 | 6.8219 |
| S_47 | 6.829 |
| S_48 | **7.1884 (-1.884 e)** |
| S_49 | 6.2587 |
| S_50 | 6.3462 |

| Atom number | Bader electron Charge partition |
|---|---|
| S_51 | 6.399 |
| S_52 | 6.3202 |
| S_53 | 7.012 |
| S_54 | 6.7907 |
| S_55 | 6.9717 |
| S_56 | 6.8171 |
| S_57 | 6.7525 |
| S_58 | 6.778 |
| S_59 | 6.8661 |
| S_60 | 6.8319 |

WS2-WS2 AB stacked c-terminated

| Atom number | Bader electron Charge partition |
|---|---|
| W_1 | 3.776 |
| W_2 | 3.7835 |
| W_3 | 4.028 (+1.972 e) |
| W_4 | 3.1061 (+2.8939 e) |
| W_5 | 3.3837 |
| W_6 | 3.7844 |
| W_7 | 3.1572 |
| W_8 | 3.861 |
| W_9 | 3.7764 |
| W_10 | 3.7824 |
| W_11 | 3.7439 |
| W_12 | 3.8124 |
| W_13 | 4.011 |
| W_14 | 3.1863 |
| W_15 | 3.271 |
| W_16 | 3.7953 |
| W_17 | 3.2411 |
| W_18 | 3.8144 |
| W_19 | 3.8133 |
| W_20 | 3.7385 |
| S_1 | 6.9723 |
| S_2 | 7.1269 (-1.1269 e) |
| S_3 | 6.9892 |
| S_4 | 6.7431 |
| S_5 | 6.7817 |
| S_6 | 6.2849 |
| S_7 | 6.9731 |
| S_8 | 6.7843 |
| S_9 | 6.3191 |
| S_10 | 6.99 |
| S_11 | 6.7422 |
| S_12 | 6.8866 |
| S_13 | 6.8061 |
| S_14 | 6.851 |
| S_15 | 6.9563 |
| S_16 | 6.8279 |
| S_17 | 7.0715 |
| S_18 | 6.7637 |
| S_19 | 6.262 (-0.262 e) |
| S_20 | 6.429 |
| S_21 | 6.4072 |
| S_22 | 6.2779 |
| S_23 | 6.9811 |
| S_24 | 7.1218 |
| S_25 | 6.98 |
| S_26 | 6.7545 |
| S_27 | 7.0399 |
| S_28 | 6.826 |
| S_29 | 6.7648 |
| S_30 | 6.8089 |
| S_31 | 7.0074 |
| S_32 | 6.7348 |
| S_33 | 6.9775 |
| S_34 | 7.0797 |
| S_35 | 6.751 |
| S_36 | 6.3303 |
| S_37 | 7.0178 |
| S_38 | 6.7889 |
| S_39 | 6.3761 |

| Atom_number | Bader electron Charge partition |
|---|---|
| S_40 | 6.9771 |
| S_41 | 7.1069 |
| S_42 | 6.7671 |
| S_43 | 6.8052 |
| S_44 | 6.8094 |
| S_45 | 6.8459 |
| S_46 | 6.8256 |
| S_47 | 6.9622 |
| S_48 | 6.7488 |
| S_49 | 6.3112 |
| S_50 | 6.4066 |
| S_51 | 6.3402 |
| S_52 | 6.392 |
| S_53 | 6.9823 |
| S_54 | 6.7491 |
| S_55 | 6.9809 |
| S_56 | 6.972 |
| S_57 | 7.1025 |
| S_58 | 6.8249 |
| S_59 | 6.7743 |
| S_60 | 6.8637 |

WS2-WSe2 AA stacked c-terminated

| Atom_number | Bader electron Charge partition |
|---|---|
| W_1 | 3.797 |
| W_2 | 3.7474 |
| W_3 | 3.9959 |
| W_4 | 3.5276 (+2.4724 e) |
| W_5 | 3.3683 |
| W_6 | 3.8111 |
| W_7 | 3.5447 |
| W_8 | 3.7824 |
| W_9 | 3.7726 |
| W_10 | 3.7833 |
| W_11 | 4.3316 |
| W_12 | 4.3383 |
| W_13 | 4.5441 (+1.4559) |
| W_14 | 3.9669 |
| W_15 | 4.1049 |
| W_16 | 4.3398 |
| W_17 | 4.0082 |
| W_18 | 4.3071 |
| W_19 | 4.3402 |
| W_20 | 4.3256 |
| S_1 | 6.9896 |
| S_2 | 6.9105 |
| S_3 | 7.0642 |
| S_4 | 6.874 |
| S_5 | 6.9015 |
| S_6 | 6.3639 |
| S_7 | 6.977 |
| S_8 | 6.8133 |
| S_9 | 6.364 |
| S_10 | 7.0421 |
| S_11 | 6.9375 |
| S_12 | 6.8194 |
| S_13 | 6.9777 |
| S_14 | 6.8338 |
| S_15 | 6.8115 |
| S_16 | 6.7619 |
| S_17 | 6.851 |
| S_18 | 6.8159 |
| S_19 | 6.4178 |
| S_20 | 6.4134 |
| S_21 | 6.2778 |
| S_22 | 6.3209 |
| S_23 | 6.9926 |
| S_24 | 6.8791 |
| S_25 | 7.0672 (-1.0672 e) |
| S_26 | 6.8285 |
| S_27 | 6.8574 |
| S_28 | 6.8456 |
| S_29 | 6.871 |
| S_30 | 6.7766 |
| Se_1 | 6.7351 |
| Se_2 | 6.669 |
| Se_3 | 6.7035 |
| Se_4 | 6.5131 |
| Se_5 | 6.5011 |
| Se_6 | 6.3058 |
| Se_7 | 6.7287 |
| Se_8 | 6.594 |
| Se_9 | 6.284 |
| Se_10 | 6.719 |
| Se_11 | 6.5896 |
| Se_12 | 6.5702 |
| Se_13 | 6.622 |
| Se_14 | 6.525 |
| Se_15 | 6.5722 |
| Se_16 | 6.6337 |
| Se_17 | 6.524 |
| Se_18 | 6.6461 |
| Se_19 | 6.3137 |
| Se_20 | 6.2616 |
| Se_21 | 6.3028 |
| Se_22 | 6.2462 (-0.2462 e) |

| Atom_number | Bader electron Charge partition |
|---|---|
| Se_23 | 6.7441 |
| Se_24 | 6.6438 |
| Se_25 | 6.7278 |
| Se_26 | 6.5215 |
| Se_27 | 6.6308 |
| Se_28 | 6.534 |
| Se_29 | 6.5993 |
| Se_30 | 6.6446 |

WS2-WSe2 AB stacked c-terminated

| Atom_number | Bader electron Charge partition |
|---|---|
| W_1 | 3.8172 |
| W_2 | 3.8055 |
| W_3 | 4.0469 |
| W_4 | 3.0693 |
| W_5 | 3.0431 (+2.9569 e) |
| W_6 | 3.7794 |
| W_7 | 3.5063 |
| W_8 | 3.7834 |
| W_9 | 3.7883 |
| W_10 | 3.7839 |
| W_11 | 4.3183 |
| W_12 | 4.3459 |
| W_13 | 4.5561 (+1.4439 e) |
| W_14 | 4.0523 |
| W_15 | 4.074 |
| W_16 | 4.3104 |
| W_17 | 4.0825 |
| W_18 | 4.3383 |
| W_19 | 4.3219 |
| W_20 | 4.2984 |
| S_1 | 6.9737 |
| S_2 | 7.1226 |
| S_3 | 7.0423 |
| S_4 | 6.7926 |
| S_5 | 6.83 |
| S_6 | 6.333 |
| S_7 | 6.9874 |
| S_8 | 6.8127 |
| S_9 | 6.3199 |
| S_10 | 7.0374 |
| S_11 | 7.0652 |
| S_12 | 6.7993 |
| S_13 | 7.1308 |
| S_14 | 6.8391 |
| S_15 | 6.813 |
| S_16 | 6.7888 |
| S_17 | 6.8474 |
| S_18 | 6.8381 |
| S_19 | 6.3101 |
| S_20 | 6.3536 |
| S_21 | 6.4066 |
| S_22 | 6.42 |
| S_23 | 6.9631 |
| S_24 | 7.1364 (-1.1364 e) |
| S_25 | 7.0396 |
| S_26 | 6.796 |
| S_27 | 6.849 |
| S_28 | 6.8576 |
| S_29 | 6.9085 |
| S_30 | 6.8442 |
| Se_1 | 6.7079 |
| Se_2 | 6.4865 |
| Se_3 | 6.7408 |
| Se_4 | 6.6889 |
| Se_5 | 6.554 |
| Se_6 | 6.2752 |
| Se_7 | 6.7092 |
| Se_8 | 6.6152 |
| Se_9 | 6.2383 |
| Se_10 | 6.7337 |
| Se_11 | 6.6231 |

| Atom_number | Bader Charge |
|---|---|
| Se_12 | 6.5508 |
| Se_13 | 6.6329 |
| Se_14 | 6.5276 |
| Se_15 | 6.5986 |
| Se_16 | 6.6331 |
| Se_17 | 6.5621 |
| Se_18 | 6.56 |
| Se_19 | 6.2596 |
| Se_20 | 6.2189 (-0.2189 e) |
| Se_21 | 6.2993 |
| Se_22 | 6.321 |
| Se_23 | 6.7176 |
| Se_24 | 6.5026 |
| Se_25 | 6.7311 |
| Se_26 | 6.6858 |
| Se_27 | 6.6488 |
| Se_28 | 6.542 |
| Se_29 | 6.6215 |
| Se_30 | 6.6344 |

**WS$_2$-WS$_2$ AA stacked *m*-terminated**

| Atom_number | Bader Charge |
|---|---|
| W_1 | 4.5663 |
| W_2 | 4.2158 |
| W_3 | 4.2431 |
| W_4 | 4.7545 |
| W_5 | 4.3288 |
| W_6 | 4.6824 |
| W_7 | 4.4091 |
| W_8 | 4.628 |
| W_9 | 4.5508 |
| W_10 | 4.8743 |
| W_11 | 4.8599 |
| W_12 | 4.6719 |
| W_13 | 4.14 |
| W_14 | 4.4481 |
| W_15 | 4.6043 |
| W_16 | **3.3687 (+2.6313 e)** |
| W_17 | 4.2024 |
| W_18 | 4.1992 |
| W_19 | 4.8176 |
| W_20 | 4.3493 |
| W_21 | 4.5339 |
| W_22 | 4.4124 |
| W_23 | 4.7856 |
| W_24 | 4.6933 |
| W_25 | 4.9682 |
| W_26 | 4.8853 |
| W_27 | **5.8745 (+0.1255 e)** |
| W_28 | 4.1641 |
| W_29 | 4.303 |
| W_30 | 4.622 |
| S_1 | 7.0966 |
| S_2 | 7.1012 |
| S_3 | 7.1106 |
| S_4 | 7.1054 |
| S_5 | 7.1241 |
| S_6 | **6.9867 (-0.9867 e)** |
| S_7 | **7.2023 (-1.2023 e)** |
| S_8 | 7.0216 |
| S_9 | 7.1046 |
| S_10 | 7.1554 |
| S_11 | 7.0629 |
| S_12 | 7.1326 |
| S_13 | 7.1085 |
| S_14 | 7.0579 |
| S_15 | 7.0783 |
| S_16 | 7.0714 |
| S_17 | 7.0967 |
| S_18 | 7.1049 |
| S_19 | 7.1292 |
| S_20 | 7.1387 |

| Atom_number | Bader Charge |
| --- | --- |
| S_21 | 7.1225 |
| S_22 | 7.1246 |
| S_23 | 7.0965 |
| S_24 | 7.0561 |
| S_25 | 7.1245 |
| S_26 | 7.009 |
| S_27 | 7.0983 |
| S_28 | 7.011 |
| S_29 | 7.1537 |
| S_30 | 7.0837 |
| S_31 | 7.1102 |
| S_32 | 7.0801 |
| S_33 | 7.1494 |
| S_34 | 7.0445 |
| S_35 | 7.1335 |
| S_36 | 7.0984 |
| S_37 | 7.0793 |
| S_38 | 7.0923 |
| S_39 | 7.1077 |
| S_40 | 7.0784 |

WS2-WS2 AB stacked m-terminated

| Atom_number | Bader Charge |
| --- | --- |
| W_1 | 4.4886 |
| W_2 | 4.2469 |
| W_3 | 4.2146 |
| W_4 | 4.8674 |
| W_5 | 4.4671 |
| W_6 | 4.6867 |
| W_7 | 4.2574 |
| W_8 | 4.5256 |
| W_9 | 4.4837 |
| W_10 | 4.869 |
| W_11 | 4.9025 |
| W_12 | 4.7257 |
| W_13 | 4.2139 |
| W_14 | 4.3839 |
| W_15 | 4.6559 |
| W_16 | 5.8867 (+0.1133 e) |
| W_17 | 4.1813 |
| W_18 | 4.1681 |
| W_19 | 4.8505 |
| W_20 | 4.3921 |
| W_21 | 4.5962 |
| W_22 | 4.3804 |
| W_23 | 4.6607 |
| W_24 | 4.7543 |
| W_25 | 4.896 |
| W_26 | 4.8297 |
| W_27 | 3.3663 (+2.6337 e) |
| W_28 | 4.1753 |
| W_29 | 4.4569 |
| W_30 | 4.5708 |
| S_1 | 7.1052 |
| S_2 | 7.1353 |
| S_3 | 7.1525 |
| S_4 | 7.1465 (-1.1465 e) |
| S_5 | 7.083 |
| S_6 | 6.9985 |
| S_7 | 7.1324 |
| S_8 | 7.019 |
| S_9 | 7.0602 |
| S_10 | 7.0799 |
| S_11 | 7.0661 |
| S_12 | 7.1753 |
| S_13 | 7.0803 |
| S_14 | 7.0791 |
| S_15 | 7.0987 |
| S_16 | 7.112 |
| S_17 | 7.0955 |
| S_18 | 7.1075 |
| S_19 | 7.1464 |
| S_20 | 7.138 |

| Atom | Bader Charge |
|---|---|
| S_21 | 7.0561 |
| S_22 | 7.0859 |
| S_23 | 7.1043 |
| S_24 | 7.0741 |
| S_25 | 7.0517 |
| S_26 | 6.9975 (-0.9975 e) |
| S_27 | 7.1033 |
| S_28 | 7.0087 |
| S_29 | 7.1075 |
| S_30 | 7.1329 |
| S_31 | 7.1196 |
| S_32 | 7.1283 |
| S_33 | 7.1278 |
| S_34 | 7.072 |
| S_35 | 7.1308 |
| S_36 | 7.0992 |
| S_37 | 7.1386 |
| S_38 | 7.1046 |
| S_39 | 7.1049 |
| S_40 | 7.0866 |

WS2-WSe2 AA stacked m-terminated

| Atom_number | Bader Charge |
|---|---|
| W_1 | 4.6471 |
| W_2 | 4.2029 (+1.7971 e) |
| W_3 | 4.2598 |
| W_4 | 4.8404 |
| W_5 | 4.308 |
| W_6 | 4.6683 |
| W_7 | 4.4788 |
| W_8 | 4.6546 |
| W_9 | 4.3201 |
| W_10 | 4.9066 |
| W_11 | 4.7937 |
| W_12 | 4.5668 |
| W_13 | 4.1933 |
| W_14 | 4.329 |
| W_15 | 4.773 |
| W_16 | 4.9187 |
| W_17 | 4.8499 |
| W_18 | 4.762 |
| W_19 | 5.1762 (+0.8238 e) |
| W_20 | 4.8153 |
| W_21 | 4.7564 |
| W_22 | 4.7219 |
| W_23 | 4.9417 |
| W_24 | 4.871 |
| W_25 | 5.1712 |
| W_26 | 5.1122 |
| W_27 | 4.8267 |
| W_28 | 4.6797 |
| W_29 | 4.7626 |
| W_30 | 4.8875 |
| S_1 | 7.1153 |
| S_2 | 7.0836 |
| S_3 | 7.1358 |
| S_4 | 7.091 |
| S_5 | 7.0858 |
| S_6 | 7.0045 |
| S_7 | 7.1355 |
| S_8 | 7.0303 |
| S_9 | 7.0921 |
| S_10 | 7.1645 |
| S_11 | 7.0719 |
| S_12 | 7.1066 |
| S_13 | 7.0974 |
| S_14 | 7.0696 |
| S_15 | 7.1399 |
| S_16 | 7.0856 |
| S_17 | 7.1278 |
| S_18 | 7.0921 |
| S_19 | 7.1499 |
| S_20 | 7.1658 (-1.1658 e) |

| Atom_number | Bader Charge |
|---|---|
| Se_1 | 6.8588 |
| Se_2 | 6.8967 |
| Se_3 | 6.8725 |
| Se_4 | 6.8416 |
| Se_5 | 6.8907 |
| Se_6 | 6.7453 |
| Se_7 | 6.7924 |
| Se_8 | 6.756 |
| Se_9 | 6.8906 |
| Se_10 | 6.7651 |
| Se_11 | 6.814 (-0.814 e) |
| Se_12 | 6.8606 |
| Se_13 | 6.8941 |
| Se_14 | 6.8145 |
| Se_15 | 6.8262 |
| Se_16 | 6.8417 |
| Se_17 | 6.8302 |
| Se_18 | 6.8661 |
| Se_19 | 6.8744 |
| Se_20 | 6.8278 |

WS2-WSe2 AB stacked m-terminated

| Atom_number | Bader Charge |
|---|---|
| W_1 | 4.7064 |
| W_2 | 4.2549 |
| W_3 | 4.1701 |
| W_4 | 4.8452 |
| W_5 | 4.5263 |
| W_6 | 4.6113 |
| W_7 | 4.3729 |
| W_8 | 4.7298 |
| W_9 | 5.9104 (+0.0896 e) |
| W_10 | 4.848 |
| W_11 | 4.7699 |
| W_12 | 4.6159 |
| W_13 | 4.2307 |
| W_14 | 4.3277 |
| W_15 | 3.2404 (+2.7596 e) |
| W_16 | 4.8426 |
| W_17 | 4.8483 |
| W_18 | 4.6918 |
| W_19 | 5.1919 |
| W_20 | 4.5752 |
| W_21 | 4.9508 |
| W_22 | 4.81 |
| W_23 | 4.9561 |
| W_24 | 4.7622 |
| W_25 | 5.1297 |
| W_26 | 4.9477 |
| W_27 | 4.7798 |
| W_28 | 4.8141 |
| W_29 | 4.8387 |
| W_30 | 4.8014 |
| S_1 | 7.0792 |
| S_2 | 7.0448 |
| S_3 | 7.1231 |
| S_4 | 7.0686 |
| S_5 | 7.0683 |
| S_6 | 6.9849 |
| S_7 | 7.1724 (-1.1724 e) |
| S_8 | 7.0104 |
| S_9 | 7.1215 |
| S_10 | 7.1466 |
| S_11 | 7.0623 |
| S_12 | 7.1017 |
| S_13 | 7.0826 |
| S_14 | 7.0867 |
| S_15 | 7.0803 |
| S_16 | 7.1168 |
| S_17 | 7.1104 |
| S_18 | 7.1305 |
| S_19 | 7.0833 |
| S_20 | 7.1158 |
| Se_1 | 6.8252 |
| Se_2 | 6.9682 |
| Se_3 | 6.9057 |
| Se_4 | 6.8218 |
| Se_5 | 6.8117 |
| Se_6 | 6.6598 |
| Se_7 | 6.8225 |
| Se_8 | 6.7499 (-0.7499 e) |
| Se_9 | 6.9719 |
| Se_10 | 6.8909 |
| Se_11 | 6.8703 |
| Se_12 | 6.851 |
| Se_13 | 6.9173 |
| Se_14 | 6.8237 |
| Se_15 | 6.8648 |
| Se_16 | 6.872 |
| Se_17 | 6.8819 |
| Se_18 | 6.8758 |
| Se_19 | 6.8956 |
| Se_20 | 6.8299 |